\documentclass{elsart}
\usepackage{amssymb}

\renewcommand{\bar}[1]{\overline{#1}}

\usepackage{indentfirst}
\usepackage{psfig,color}
\usepackage{epsfig}
\usepackage{epsf}
\usepackage{graphicx}

\journal{Physics Letters B}

\begin{document}

\begin{frontmatter}
\title{Effect of asymmetric strange-antistrange sea to the NuTeV anomaly}

\author[pku]{Yong Ding},
\author[pku]{Rong-Guang Xu},
\author[ccastpku]{Bo-Qiang Ma\corauthref{cor}}
\corauth[cor]{Corresponding author.} \ead{mabq@phy.pku.edu.cn}
\address[pku]{Department of Physics, Peking University, Beijing 100871, China}
\address[ccastpku]{CCAST (World Laboratory), P.O.~Box 8730, Beijing
100080, China\\
Department of Physics, Peking University, Beijing 100871, China}

\begin{abstract}
We calculate the strange quark and antiquark distributions of the
nucleon by using the effective chiral quark model, and find that
the strange-antistrange asymmetry can bring a contribution of
about 60--100\% to the NuTeV deviation of $\sin^{2}\theta_{w}$
from the standard value measured in other electroweak processes.
The results are insensitive to different inputs. The light-flavor
quark asymmetry of $\bar{d}-\bar{u}$ is also investigated and
found to be consistent with the experimental measurements.
Therefore the chiral quark model provides a successful picture to
understand the NuTeV anomaly, as well as the light-flavor quark
asymmetry and the proton spin problem in previous studies.
\end{abstract}

\end{frontmatter}

\par
The nucleon sea is a very active research direction of hadron
physics due to its rich phenomena which are different from naive
theoretical expectations and intriguing to understand strong
interaction. Among various topics, the strange content of the
nucleon sea is one of the most attractive issues, due to its close
connection to the proton spin problem~\cite{bek88} and to the
obscure situation about the strange-antistrange
asymmetry~\cite{bm97}. Although much progress and achievement have
been made both theoretically and experimentally, our knowledge of
the strange sea is still limited. A common assumption about the
strange sea is that the $s$ and $\bar{s}$ distributions are
symmetric, but in fact this is established neither theoretically
nor experimentally. Possible manifestations of nonperturbative
effects for the strange-antistrange asymmetry have been discussed
along with some phenomenological
explanations~\cite{bm97,st87,bw92,hss96,cm98,cs99}. Also there
have been some experimental analyses~\cite{b95,sbr97,a97,bpz00},
which suggest the $s$-$\bar{s}$ asymmetry of the nucleon sea.
Therefore, the precision measurement of strange quark and
antiquark distributions in the nucleon is one of the challenging
and significant tasks for experimental physics.

The NuTeV Collaboration~\cite{zell02} reported the value of
$\sin^{2}\theta_{w}$ measured in deep inelastic scattering~(DIS)
on nuclear target with both neutrino and antineutrino beams.
Having considered and examined various source of systematic
errors, the NuTeV Collaboration had the value:
$$\sin^{2}\theta_{w}=0.2277\pm0.0013~(\mbox{stat})\pm0.0009~(\mbox{syst}),$$
which is three standard deviations larger than the value
$\sin^{2}\theta_{w}=0.2227\pm0.0004$ measured in other electroweak
processes, where $\theta_{w}$ is the Weinberg angle which is one
of the important quantities in the standard model. The NuTeV
Collaboration measured the value of $\sin^{2}\theta_{w}$ by using
the ratio of neutrino neutral-current and charged-current cross
sections on iron~\cite{zell02}. This procedure is closely related
to the Paschos-Wolfenstein~(P-W) relation~\cite{pash73}:
\begin{equation}
R^{-}=\frac{\sigma^{{\nu}N}_{NC}-\sigma^{\overline{\nu}N}_{NC}}{\sigma^{{\nu}N}_{CC}
-\sigma^{\overline{\nu}N}_{CC}}=\frac{1}{2}-\sin^{2}\theta_{w},
\label{ratio}
\end{equation}
which is based on the assumptions of charge symmetry, isoscalar
target and $s(x)=\bar{s}(x)$. There have been a number of
corrections considered for the P-W relation, for example: charge
symmetry violation~\cite{lt03}, neutron excess~\cite{k02}, nuclear
effect~\cite{ksy02}, strange-antistrange
asymmetry~\cite{cs03,dm04}, and also source for physics beyond
standard model~\cite{d02}. It is still obscure whether the
strange-antistrange asymmetry can account for this NuTeV
anomaly~\cite{o03}. Cao and Signal~\cite{cs03} reexamined the
strange-antistrange asymmetry using the meson cloud model and
concluded that the second moment $S^{-}\equiv\int^{1}_{0}
x[s(x)-\bar{s}(x)]\textmd{d}x$ is fairly small and unlikely to
affect the NuTeV extraction of $\sin^{2}\theta_{w}$. Oppositely,
Brodsky and Ma~\cite{bm97} proposed a light-cone meson-baryon
fluctuation model to describe the $s(x)-\bar{s}(x)$ distributions
and found a significantly different case from what obtained by
using the meson cloud model~\cite{st87,hss96}, as has been
illustrated recently~\cite{dm04}. Also, Szczurek {\it et
al.}~\cite{sbf96} suggested that the effect of SU(3)$_{f}$
symmetry violation may be specially important in understanding the
strangeness content of the nucleon within the effective chiral
quark model, and compared their results with those of the
traditional meson cloud model qualitatively. In this letter, we
focus our attention on the distributions of $s(x)$ and
$\bar{s}(x)$, and calculate the second moment $S^{-}$ by using the
effective chiral quark model. We find that the $s$-$\bar{s}$
asymmetry can remove the NuTeV anomaly by about 60--100\%, and
that the results are insensitive to different inputs.

The effective chiral quark model~\cite{mg84}, which was formulated
by Manohar and Georgi, is successful in explaining the Gottfried
sum rule violation reported by the New Muon
Collaboration~\cite{nmc91}, first done by Eichten, Hinchliffe and
Quigg~\cite{ehq92}. This model also plays an important role in
explaining the proton spin problem~\cite{a89} by Cheng and
Li~\cite{cl95}. These successes naturally lead us to study the
strange quark and antiquark distributions and confront them with
the NuTeV result within the effective chiral quark picture. In the
effective chiral quark model, the relevant degrees of freedom are
constituent quarks, gluons and Goldstone~(GS) bosons. It is
noticeable that the effect of the internal gluon is small, when
compared with those of the GS bosons and quarks, so it is
negligible in this work. In this picture, the constituent quarks
couple directly to the GS bosons, which are the consequences of
the spontaneously broken chiral symmetry, and any low energy
hadron properties should include this symmetry violation. The
effective interaction Lagrangian is
\begin{equation}
L=\bar{\psi}(iD_{\mu}+V_{\mu})\gamma^{\mu}\psi+ig_{A}\bar{\psi}A_{\mu}\gamma^{\mu}\gamma_{5}\psi+\cdots,
\end{equation}
where
\begin{equation}
\psi=\left(%
\begin{array}{c}
  u \\
  d \\
  s \\
\end{array}%
\right)
\end{equation}
is the quark field and $D_{\mu}$ is the covariant derivative. The
vector~($V_{\mu}$) and axial-vector~($A_{\mu}$) currents are
defined in terms of GS bosons:
\begin{equation}
\left(%
\begin{array}{c}
  V_{\mu} \\
  A_{\mu} \\
\end{array}%
\right)=\frac{1}{2}(\xi^{+}\partial_{\mu}\xi\pm\xi\partial_{\mu}\xi^{+}),
\end{equation}
where $\xi=\mathrm{exp}(i\Pi/f)$ and $\Pi$ has the form:
\begin{equation}
\Pi\equiv\frac{1}{\sqrt{2}}\left(
\begin{array}{ccc}
  \frac{\pi^{0}}{\sqrt{2}}+\frac{\eta}{\sqrt{6}} & \pi^{+} & K^{+} \\
  \pi^{-} & -\frac{\pi^{0}}{\sqrt{2}}+\frac{\eta}{\sqrt{6}} & K^{0} \\
  K^{-} & \bar{K^{0}} & \frac{-2\eta}{\sqrt{6}} \\
\end{array}
\right).
\end{equation}
Expanding $V_{\mu}$ and $A_{\mu}$ in power of $\Pi/f$ gives
$V_{\mu}=0+O(\Pi/f)^{2}$ and
$A_{\mu}=i\partial_{\mu}\Pi/f+O(\Pi/f)^{2}$, where the
pseudoscalar decay constant is $f\simeq93$~MeV. So the effective
interaction between GS bosons and quarks becomes~\cite{ehq92}
\begin{equation}
L_{\Pi
q}=-\frac{g_{A}}{f}\bar{\psi}\partial_{\mu}\Pi\gamma^{\mu}\gamma_{5}\psi.
\end{equation}
The framework that we use is based on timed-ordered perturbative
theory in the infinite momentum frame~(IMF), in which all
particles are on-mass-shell so that the factorization of
subprocess is automatic. We can express the quark distributions
inside a nucleon as a convolution of a constituent quark
distribution in a nucleon and the structure of a constituent
quark. The light-front Fock decompositions of constituent quark
wave functions have
\begin{eqnarray}
|U\rangle&=&Z^{\frac{1}{2}}|u_{0}\rangle+a_{\pi}|u\pi^{0}\rangle\nonumber\\&&
+\frac{a_{\pi}}{\sqrt{2}}|d\pi^{+}\rangle+
a_{K}|sK^{+}\rangle+\frac{a_{\eta}}{\sqrt{6}}|u\eta\rangle,\\
|D\rangle&=&Z^{\frac{1}{2}}|d_{0}\rangle+a_{\pi}|d\pi^{0}\rangle\nonumber\\&&
+\frac{a_{\pi}}{\sqrt{2}}|u\pi^{-}\rangle+
a_{K}|sK^{0}\rangle+\frac{a_{\eta}}{\sqrt{6}}|d\eta\rangle,
\end{eqnarray}
where $Z$ is the renormalization constant for the bare constituent
quark and $|a_{\alpha}|^{2}$ are the probabilities to find GS
bosons in the dressed constituent quark states $|U\rangle$ for an
$up$ quark and $|D\rangle$ for a $down$ quark. In chiral field
theory, the spin-independent term is given by~\cite{sw98}
\begin{equation}
q_{j}(x)=\int^{1}_{0}\frac{\textmd{d}y}{y}P_{j\alpha/i}(y)q_{i}(\frac{x}{y}).
\end{equation}
Here, $P_{j\alpha/i}(y)$ is the splitting function which gives the
probability for finding a constituent quark $j$ carrying the the
light-cone momentum fraction $y$ together with a spectator GS
boson~($\alpha=\pi, K, \eta$), both of which coming from a parent
constituent quark $i$:
\begin{eqnarray}
P_{j\alpha/i}(y)=\frac{1}{8\pi^{2}}(\frac{g_{A}\bar{m}}{f})^{2}\int
\textmd{d}k^{2}_{T}\frac{(m_{j}-m_{i}y)^{2}+k^{2}_{T}}{y^{2}(1-y)[m_{i}^{2}-M^{2}_{j\alpha}]^{2}},
\nonumber
\end{eqnarray}
where $m_{i}, m_{j}, m_{\alpha}$ are the masses of the $i,
j$-constituent quarks and the pseudosclar meson $\alpha$,
respectively,
\begin{equation}
M^{2}_{j\alpha}=\frac{m^{2}_{j}+k^{2}_{T}}{y}+\frac{m^{2}_{\alpha}+k^{2}_{T}}{1-y}
\end{equation}
is the invariant mass squared of the final state, and
$\bar{m}=(m_{i}+m_{j})/2$ is the average mass of the constituent
quarks.  We choose $m_{u}=m_{d}=330$~MeV, $m_{s}=480$~MeV,
$m_{\pi^{\pm}}=m_{\pi^{0}}=140$~MeV and
$m_{K^{+}}=m_{K^{0}}=495$~MeV. We adopt the definition of the
first moment of splitting function: $\langle
P_{j\alpha/i}\rangle=\int^{1}_{0}P_{j\alpha/i}(x)\textmd{d}x$ and
$\langle P_{j\alpha/i}\rangle=\langle P_{\alpha j/i}\rangle\equiv
\langle P_{\alpha}\rangle=|a_{\alpha}|^{2}$~\cite{sw98}. It is
conventional that an exponential cutoff is used in IMF
calculations. Usually
\begin{equation}
g_{A}=g_{A}^{\prime}\textmd{exp}\bigg{[}\frac{m^{2}_{i}-M^{2}_{j\alpha}}{4\Lambda^{2}}\bigg{]},
\end{equation}
with $g_{A}^{\prime}=1$ following the large $N_{c}$
argument~\cite{w90}, $\Lambda$ is the cutoff parameter, which is
determined by the experiment data of the Gottfried sum and the
constituent mass input for $\pi$, but for $K$ and $\eta$, the
terms $\langle P_{K}\rangle$ and $\langle P_{\eta}\rangle$ in the
Gottfried sum  cancel with those in $Z=1-\frac{3}{2}\langle
P_{\pi}\rangle-\langle P_{K}\rangle-\frac{1}{6}\langle
P_{\eta}\rangle$:
\begin{eqnarray}
S_{\mathrm{Gottfried}}&=&\int^{1}_{0}\frac{\mathrm{d}x}{x}[F^{p}_{2}(x)-F^{n}_{2}(x)]\nonumber\\&
=&\frac{1}{3}(Z-\frac{1}{2}\left<P_{\pi}\right>+\left<P_{K}\right>+\frac{1}{6}\left<P_{\eta}\right>)\nonumber\\&
=&\frac{1}{3}(1-2\left<P_{\pi}\right>).
\end{eqnarray}
Usually $\Lambda_{K}$ was given by
$\Lambda_{K}=\Lambda_{\pi}=1500$~MeV~\cite{sbf96,sw98}, however,
the $SU_{f}(3)$ symmetry breaking requires smaller $\langle
P_{K}\rangle$ and $\langle P_{\eta}\rangle$ \cite{song}, so that
we should adopt a smaller value for $\Lambda_{K}$ such as from
900~MeV to 1100~MeV.

When probing the internal structure of the GS bosons, the process
can be written in the following form~\cite{sw98}:
\begin{equation}
q_{k}(x)=\int\frac{\textmd{d}y_{1}}{y_{1}}\frac{\mathrm d
y_{2}}{y_{2}}V_{k/\alpha}(\frac{x}{y_{1}})P_{\alpha
j/i}(\frac{y_{1}}{y_{2}})q_{i}(y_{2}),
\end{equation}
where $P_{\alpha j/i}(x)=P_{j\alpha/i}(1-x)$ and $V_{k/\alpha}(x)$
is the quark $k$ distribution function in $\alpha$ and is
normalized to 1. Because the mass of $\eta$ is so high and the
coefficient is so small that the fluctuation of it is suppressed,
the contribution is not considered here. Assuming that the bare
quark distribution functions are given in terms of the constituent
quark distributions $u_{0}$ and $d_{0}$, which are normalized, we
have:
\begin{eqnarray}
u(x)&=&Zu_{0}(x)+P_{u\pi^{-}/d}\otimes d_{0}+V_{u/\pi^{+}}\otimes
P_{\pi^{+}d/u}\otimes
u_{0}\nonumber\\&+&\frac{1}{2}P_{u\pi^{0}/u}\otimes u_{0}+V_{u/K^{+}}\otimes P_{K^{+}s/u}\otimes u_{0}\nonumber\\
&+&\frac{1}{4}V_{u/\pi^{0}}\otimes (P_{\pi^{0}u/u}\otimes u_{0}+P_{\pi^{0}d/d}\otimes d_{0}),\nonumber\\
d(x)&=&Zd_{0}(x)+P_{d\pi^{+}/u}\otimes u_{0}+V_{d/\pi^{-}}\otimes
P_{\pi^{-}u/d }\otimes d_{0}\nonumber\\&+&
\frac{1}{2}P_{d\pi^{0}/d}\otimes d_{0}+V_{d/K^{0}}\otimes
P_{K^{0}s/d}\otimes d_{0} \nonumber
\\&+&\frac{1}{4}V_{d/\pi^{0}}\otimes (P_{\pi^{0}u/u }\otimes
u_{0}+P_{\pi^{0}d/d}\otimes d_{0}).\nonumber
\end{eqnarray}
Here, we define the notation for the convolution integral:
\begin{equation}
P\otimes q=\int_{x}^{1}\frac{\textmd{d}y}{y}P(y)q(\frac{x}{y}).
\end{equation}
In the same way, we can have the light-flavor antiquark and
strange quark and antiquark distributions:
\begin{eqnarray}
\bar{u}(x)&=&V_{\bar{u}/\pi^{-}}\otimes P_{\pi^{-}u/d}\otimes
d_{0}\nonumber\\&+&\frac{1}{4}V_{\bar{u}/\pi^{0}}\otimes
(P_{\pi^{0}u/u}\otimes
u_{0}+P_{\pi^{0}d/d}\otimes d_{0}),\nonumber\\
\bar{d}(x)&=&V_{\bar{d}/\pi^{+}}\otimes P_{\pi^{+}d/u}\otimes
u_{0}\nonumber\\&+&\frac{1}{4}V_{\bar{d}/\pi^{0}}\otimes
(P_{\pi^{0}u/u}\otimes
u_{0}+P_{\pi^{0}d/d}\otimes d_{0}),\nonumber\\
 s(x)&=&P_{sK^{+}/u}\otimes u_{0}+P_{sK^{0}/d}\otimes d_{0},\nonumber\\
\bar{s}(x)&=&V_{\bar{s}/K^{+}}\otimes P_{K^{+}s/u}\otimes
 u_{0}+V_{\bar{s}/K^{0}}\otimes P_{K^{0}s/d}\otimes d_{0},~~\nonumber
\end{eqnarray}
where
$V_{u/\pi^{+}}=V_{\bar{d}/\pi^{+}}=V_{d/\pi^{-}}=V_{\bar{u}/\pi^{-}}=2V_{u/\pi^{0}}
=2V_{\bar{u}/\pi^{0}}=2V_{d/\pi^{0}}=2V_{\bar{d}/\pi^{0}}$,
$V_{\bar{s}/K^{+}}=V_{\bar{s}/K^{0}}$ and
$V_{u/K^{+}}=V_{d/K^{0}}$ are taken from GRS98 parametrization of
parton distributions for mesons~\cite{grs98}. The valence
distributions $u_{v}(x)=u(x)-\bar{u}(x)$ and
$d_{v}(x)=d(x)-\bar{d}(x)$ are examined to satisfy the correction
normalization with the renormalization constant $Z$. From above
procedure, we can calculate $S^{-}\equiv\int^{1}_{0}
x[s(x)-\bar{s}(x)]\textmd{d}x$, which can bring the correction in
the modified P-W relation~\cite{dm04}
\begin{equation}
R^{-}_{N}=\frac{\sigma^{\nu
N}_{NC}-\sigma^{\bar{\nu}N}_{NC}}{\sigma^{\nu
N}_{CC}-\sigma^{\bar{\nu}N}_{CC}} = R^{-}-\delta
R^{-}_{s},\label{correction}
\end{equation}
where $\delta R^{-}_{s}$ is the correction term to the P-W
relation, which comes from the asymmetry of strangeness and reads:
\begin{equation}
\delta
R^{-}_{s}=(1-\frac{7}{3}\sin^{2}\theta_{w})\frac{S^{-}}{Q_v+3
S^{-}},\label{rs}
\end{equation}
where $Q_v \equiv\int^{1}_{0} x[u_{v}(x)+d_{v}(x)]\textmd{d} x$.
Thus what measured by NuTeV should be $\sin^{2}\theta_{w}+\delta
R^{-}_{s}$, rather than $\sin^{2}\theta_{w}$ from a strict sense.
One would need $\delta R^{-}_{s}\approx 0.005$ to completely
explain the NuTeV deviation from the standard value of
$\sin^{2}\theta_{w}$ measured in other processes.

We choose two different sets of constituent quark distributions as
inputs: constituent quark (CQ) model distributions~\cite{hz81} and
CTEQ6 parametrization~\cite{p02}. The constituent quark~(CQ) model
distributions have the form with the initial scale
$Q^{2}_{0}=0.4$~Ge$\textmd{V}^{2}$:
\begin{eqnarray}
u_{0}(x)&=&\frac{2}{\textmd{B}[c_{1}+1,c_{1}+c_{2}+2]}x^{c_{1}}(1-x)^{c_{1}+c_{2}+1},
\nonumber\\
d_{0}(x)&=&\frac{1}{\textmd{B}[c_{2}+1,2c_{1}+2]}x^{c_{2}}(1-x)^{2c_{1}+1},
\end{eqnarray}
which is independent of nature of probe and its $Q^{2}$ value.
Where $B[i,j]$ is the Euler beta function with $c_{1}=0.65$ and
$c_{2}=0.35$ given in~\cite{hz81}. The other input we adopted is
from CTEQ6 parametrization with $Q_{0}=1.3$~GeV:
\begin{eqnarray}
u_{0}(x)&=&1.7199x^{-0.4474}(1-x)^{2.9009}\cdot\nonumber\\&&\textmd{
exp}[-2.3502x](1+\textmd{exp}[1.6123]x)^{1.5917},~~\nonumber\\
d_{0}(x)&=&1.4473x^{-0.3840}(1-x)^{4.9670}\cdot\nonumber\\&&
\textmd{exp}[-0.8408x](1+\textmd{
exp}[0.4031]x)^{3.0000}.~~\label{cteq}
\end{eqnarray}

\begin{figure}
\begin{center}
\scalebox{0.78}{\includegraphics[0,16][295,212]{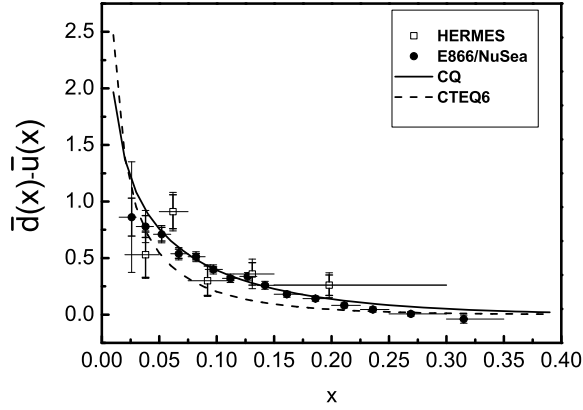}}
\caption{\small Distributions for $\bar{d}(x)-\bar{u}(x)$ for
$\Lambda_{\pi}=1500$~MeV, the solid curve for constituent quark
(CQ) model as input and the dashed curve for CTEQ6 parametrization
as input within the chiral quark model. The data are from
HERMES~($Q^{2}=2.3$~Ge$\textmd{V}^{2}/c^{2}$) and
E866/NuSea~($Q^{2}=54$~Ge$\textmd{V}^{2}/c^{2}$)
experiments~\cite{h98,e01}. }\label{dubar}
\end{center}
\end{figure}

\begin{center}
\begin{table}
\caption {\label{value}The calculated results for different
inputs}
\begin{center}
\begin{tabular}{|c|c|c|c|c|c|}
\hline
Parameter &\multicolumn{4}{|c|}{$\Lambda_{K}=1100$~MeV}&$\Lambda_{K}=900$~MeV\\
       \hline
Quantity       & $Z$ & $Q_{v}$ & $S^{-}$ & $\delta R_{s}^{-}$& $\delta R_{s}^{-}$ \\
  \hline
    CQ       & 0.731 & 0.846 & 0.00879 & 0.00473 & 0.00297\\
    \hline
    CTEQ6     & 0.731 & 0.362 & 0.00398 & 0.00498 & 0.00312\\
    \hline
\end{tabular}
\end{center}
\end{table}
\end{center}

\begin{figure}
\begin{center}
\scalebox{0.78}{\includegraphics[0,16][307,212]{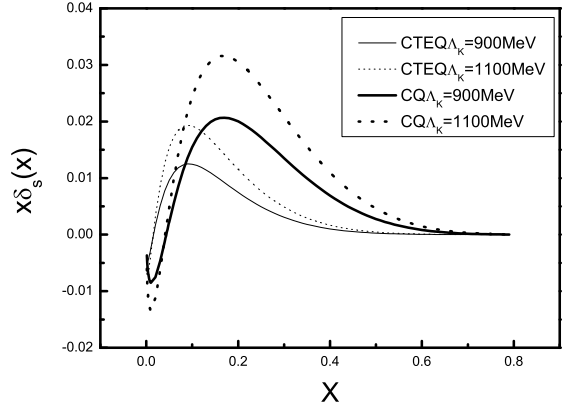}}
\caption{\small Distributions of $x\delta_{s}(x)$, with
$\delta_{s}(x)$=$s(x)-\bar{s}(x)$ for both constituent quark (CQ)
model~(thick curves) and CTEQ6 parametrization~(thin curves) as
inputs with $\Lambda_{K}=900$~MeV (solid curves) and 1100~MeV
(dashed curves ).}\label{ssbar}
\end{center}
\end{figure}

The calculated results of $\bar{d}(x)-\bar{u}(x)$ are shown in
Fig.~\ref{dubar}, from which we find that our results match the
experiments~\cite{h98,e01} well with two very different inputs of
constituent quark distributions. We also get different
distributions for $x\delta_{s}(x)$ in Fig.~\ref{ssbar}, from which
we find that the magnitudes with CQ input are almost twice larger
than those with CTEQ6 input. However, the values of $\delta
R_{s}^{-}$ in Table~\ref{value} are similar and insensitive to
different inputs at fixed $\Lambda_{K}$, as the uncertainties as
well as $Q^2$ evolution of $S^-$ and $Q_v$ in the numerator and
denominator of Eq.~(\ref{rs}) can at least partially cancel each
other. This means that the strange-antistrange asymmetry within
the framework of the effective chiral quark model can account for
about 60--100\% (corresponding to $\Lambda_{K}=900$--$1100$~MeV)
of the NuTeV anomaly without sensitivity to different inputs of
constituent quark distributions.
The adoption of a larger $\Lambda_{K}$ will bring more significant
correction to the P-W relation.

In summary, we calculated $\bar{d}(x)-\bar{u}(x)$ in the chiral
quark model with different inputs and found that the calculated
results are consistent with experiments. We also calculated
$x\delta_{s}(x)$ and found that the magnitudes are sensitive to
different inputs and parameters. However, the effect due to the
strange-antistrange asymmetry can bring a significant contribution
to the NuTeV deviation from the standard value of
$\sin^{2}\theta_{w}$, of about 60--100\% with reasonable
parameters without sensitivity to different inputs of constituent
quark distributions. Therefore the chiral quark model provides a
successful picture to understand a number of anomalies concerning
the nucleon sea: the light-flavor quark asymmetry~\cite{ehq92},
the proton spin problem~\cite{cl95}, and also the NuTeV anomaly.
This may imply that the NuTeV anomaly can be considered as a
phenomenological support to the strange-antistrange asymmetry of
the nucleon sea. Thus it is important to make a precision
measurement of the distributions of $s(x)$ and $\bar{s}(x)$ in the
nucleon more carefully in future experiments.

This work is partially supported by National Natural Science
Foundation of China.


\end{document}